\documentclass[preprint,showpacs,preprintnumbers,amsmath,amssymb]{revtex4}

\usepackage{graphicx}
\usepackage{dcolumn}
\usepackage{bm}

\begin{document}

\preprint{}

\title{Sol-gel obtained YSZ nanocrystals on a substrate: a Monte Carlo approach of the crystallographic orientation}

\author{ F.Lallet$^1$ \footnote{francois.lallet@laposte.net},N.Olivi-Tran$^{1,2}$ \footnote{olivi-tran@ges.univ-montp2.fr}}

\affiliation{$^1$Laboratoire de Sciences des Proc\'ed\'es C\'eramiques et Traitements de Surface, UMR-CNRS 6638, Ecole Nationale Sup\'erieure de C\'eramiques Industrielles, 47 avenue Albert Thomas, 87065 Limoges cedex, France;\\ $^2$Groupe d'Etude des Semi-conducteurs, UMR CNRS 5650, Universit\'e Montpellier II, Case Postale 074, Place Eug\`ene Bataillon, 34095 Montpellier cedex5, France}

\date{today}

\pagebreak

\begin{abstract}
Sol gel obtained YZS (Yttria Stabilized Zirconia) nanocrystals have different morphologies when
they grow on an $\alpha$-alumina substrate after thermal treatment. When the substrate has planar
defects, the nanocrystals grow in height and are narrow with a crystallographic orientation [111]
in the vertical direction, while when the substrate is a perfect plane at the nanometric scale, the
nanocrystals are rather extended over the substrate and do not grow in height, with a crystallo-
graphic orientation [100] in the vertical direction. We present here a Monte Carlo approach which
computes the actions of the substrate on the nanocrystals during thermal treatment: one action
is the change in crystallographic orientation depending on the presence of defects and the other is
the action on the morphology of the nanocrystals. The equivalent of thermal treatment is obtained
after applying the Metropolis algorithm with adequate expressions of the energy depending on the
inter-plane spacings and surface diffusion. Our numerical approach is in good agreement with the 
experimental results on the orientations and morphologies of the YSZ nanocrystals growing on a 
$\alpha$-alumina substrate with planar defects \cite{bachelet}.
\end{abstract}

\vfill
\maketitle

\section{Introduction}
In recent years, the formation of mesoscopic structures on crystal surfaces has become a
subject of intense experimental investigations. Generally, for non periodically ordered nanostructures, 
the increasing specific area is favorable in order to enhance the physical properties (in optics,
semiconducting etc) owing to the increased number of active sites \cite{shalaev}. \\
\indent
We will study here the evolution of a thin YSZ film deposited on a single $\alpha$-alumina
crystalline substrate with planar defects \cite{bachelet}. The thin film itself is polycrystalline with the
size of crystals corresponding to the thin film thickness. The experimental method employed
to obtain such thin films is sol gel processing \cite{brinker}. The sol gel method proceeds as follows:
a thin film is deposited at room temperature on a substrate by dip-coating. After a first
heat treatment (stage I), the thin film of nanometric thickness is made of a large amount
of nanocrystals of random orientation. At this stage the film thickness is much larger than
the mean size of these nanocrystals. After a second heat treatment at higher temperature
(stage II), thermal annealing induces grain growth. At this stage, the size of the crystals is
of the order of the film thickness. Simultaneously, the film is submitted to fragmentation
into more or less interconnected islands in order to reduce the total energy and hence to
reach a more stable state \cite{miller}. \\
\indent
The aim of this article is to model the spontaneous formation of nanoislands and
especially the evolution of their crystallographic orientations, without
matter deposition, during thermal annealing of polycrystalline nanometric thin films. 
Here, we used a Monte Carlo method applied to a polycrystalline thin film. 
This model is derived from the Solid on Solid model but is applied here in the absence of deposition. 
Our model is also derived from the two dimensional models of polycrystalline materials which computed 
the evolution of polycrystalline domains during thermal treatments \cite{srolo1, srolo2, srolo3}. 
Our model is based on energetic considerations: we compute the energies resulting from the elastic strains 
due to surface morphology of the thin film, the lattice mismatches and the grain boundary energies. 
We will see that the resulting shapes of the islands depend on the relative values of these three energies. 
This model seems to be a good approach to describe the fundamental mechanisms of nanoislands formation and
crystallographic orientation, without matter deposition, from the breaking of a thin film
under thermal annealing. \\
\indent
In section II, we shall present the model. In section III, numerical results are discussed.
Section IV corresponds to the conclusion.

\section{Numerical procedure}
We modelled a polycrystalline thin film deposited on a single crystalline substrate with
a random distribution of defects. The thin film is divided into 10$^4$ domains. Each domain
contains approximately 500 to 1000 atoms. The substrate has a random distribution of
planar defects and the domains have a section depending on the distribution of defects. The
locations of the defects are generated by a random distribution of points on the substrate.
The horizontal sections of the domains correspond to the Voronoi array of the locations of the
defects. The locations of the domains do not change during computation: no displacements
of the defects are occurring in the substrate. This model represents a polycrystalline thin
film deposited on a single crystalline substrate with planar defects. \\
\indent
In a previous article \cite{olivi}, we used the following energetic equation for the evolution of the
height of the nanocrystals:

\begin{eqnarray}
\Delta E_i & = & Al_i^2\sum_{j=1}^{NN}(h_i-h_j) \\
A & = & Y(1+\nu)\sqrt{\left( \frac{D_S\gamma_S\Delta t}{k_BT} \right)}
\end{eqnarray}

\noindent
where $h_i$ (respectively $h_j$) is the height of elementary domain $i$ (respectively $j$) and $NN$ the number of
next nearest neighbours. $Y$ is the Young modulus, $\nu$ the Poisson ratio related to the displacement $l_i$ in the horizontal direction.
$D_S$ is the self-surface diffusion constant, $\gamma_S$ the surface tension, $\Delta t$ the time interval between
two Monte Carlo steps, $k_B$ the Boltzmann constant and $T$ the absolute temperature. \\
\indent
The crystallographic reorientation of a domain may be related to the modification of the
inter-plane spacing. Indeed, for our system of $N$ lattice domains, the energy necessary to
change crystallographic orientation for domain $i$ with respect to domain $j$ becomes \cite{olivi}:

\begin{eqnarray}
\Delta E'_i & = & B\left( \frac{l_i^2}{h_i}+l_i \right)\sum_{j=1}^{NN}(d_{hi}-d_{hs}-(d_{hj}-d'_{hs})) \\
& + & C\left( \frac{l_i^2}{h_i}+l_i \right)\sum_{j=1}^{NN}(d_{vi}-d_{vs}-(d_{vj}-d'_{vs}))
\end{eqnarray}

\noindent
where $B=C=2\gamma_{jg}$ is the surface tension at the grain boundary induced by the association
of two free surfaces. The parameter $d$ is the inter-plane spacing of the family planes, thus
the value of d is defined as: $d = 1/|\mathbf{r}|$, where $\mathbf{r}$ (the reciprocal space lattice vector in m$^{-1}$),
is orthogonal to the family planes. As a consequence, $d_{vi}$ and $d_{vj}$ (respectively $d_{vs}$ and $d'_{vs}$)
are the inter-planes spacings of the family planes of the domains $i$ and $j$ (respectively of the
substrate) in the vertical plane (orthogonal to the interface). The parameters $d_{hi}$ and $d_{hj}$
(respectively $d_{hs}$ and $d'_{hs}$) are the inter-planes spacings of the family planes of the domains
$i$ and $j$ (respectively of the substrate) in the horizontal plane (parallel to the interface). See
figure 1 for a schematic view of all inter-plane spacings. In our model, each domain owns
three states ($h_i$, $d_{vi}$, $d_{hi}$); $h_i$ has its value ranging from 0 nm to a value that depends on the
physical properties of the thin film as will be seen in the results. \\
\indent
Straightforwardly, we consider here three aspects which contribute to the energy of our
thin film consisting of crystal species: the grain boundary energy (which is here equivalent
to the interfacial energy between two elementary domains of different crystallographic ori-
entations), the interfacial energy (which corresponds to the difference of energy between one
elementary domain and the substrate) and the surface energy (which is related here to the
height of each elementary domain). \\
\indent
At the beginning of computation, i.e. at $t=0$ MCS, all elementary domains were assumed
to have a height of 1 nm (the thin film is perfectly flat) and a random inter-plane spacing i.e.
0.5, 1, 1.5 or 2.5 \AA \space for both vertical and horizontal directions. The inter-plane spacings of the
substrate are chosen to be constant in the horizontal direction whereas they are as large as
the number of next nearest neighbours of the corresponding elementary domain is high. This
model is suitable to take into account the planar defects of the substrate which modifies the
inter-plane spacing of the substrate itself at its surface in the vertical direction. The Monte
Carlo algorithm works according to the classical Metropolis scheme \cite{metro}. A lattice domain is
chosen proportionally to its number of very next neighbours for three events (changes in the
projections of the crystallographic orientation and height exchange) occurring. A neighbour
of each domain is also chosen at random, and the energies given by equations (1), (2), (3)
and (4) are computed. The following numerical results have been averaged over 10 runs.

\section{Results and Discussion}
We computed all the results presented in the following figures at a temperature $T=1800$ K 
which corresponds to experimental data ($\sim$1500\textsuperscript{o}C). We have chosen $Y=300$ GPa,
$\nu=0.3$ and $D_S=8.10^{-5}$ m$^2$.s$^{-1}$\cite{ridder}; $\gamma_S=620.10^{-3}$ J.m$^{-2}$\cite{pascal}. 
$\Delta t$ is chosen to simulate the experimental time (i.e. 15 min) according to the total number of Monte-Carlo Steps 
- MCS - (i.e. 10$^9$) thus $\Delta t=9.10^{-7}$ s and $k_B\sim1.38.10^{-23}$ J.K$^{-1}$; as a consequence $A=1755.10^{13}$ J.m$^{-3}$.
According to the cubic symmetry of the YSZ phase, the surface tension at the grain
boundary can be considered as independent of the crystallographic orientation at first
order. Indeed we found in the literature a relative difference of $\sim$20$\%$ for $\gamma_{jg}$ for all
crystallographic orientations [11]. Finally we have chosen $B=C=2\gamma_jg=2\gamma_s=1240.10^{-3}$ J.m$^{-2}$. \\
\\
\indent
The aim of this section is to understand the influence of the substrate on the evolution of
the heights and crystallographic orientations of the elementary domains. We performed the
Monte Carlo process as written in section II, for a substrate with planar defects. \\
\\
\indent
We focus hereafter on the evolution of the heights of the domains. \\
In figure 2 one can see a top view of a thin film after 10$^9$ MCS. The domains are represented
by squares and the substrate appears in white; the grey level of each domain is as high as its height is small. 
This figure clearly shows the islanding process of the thin film which breaks into discrete domains during 
the computation process. This result is also illustrated on figure 3 where the number of domains of a given
height (1, 2, 3 or 4 nm) is represented as a function of the computation time (MCS). One
can note a huge increasement of the number of domain with height $h=0$ nm correlated
to the decreasement of the number of domains with height $h=1$ nm after 10$^3$ MCS. This
evolution illustrates the result presented on figure 2. One can note moreover that the number
of domains with heights 2, 3 and 4 nm increase during a short time and vanish after. From
figure 2 it is clear that the islanding process is very efficient; as a consequence, the height
of the domains increase dramatically higher than 2, 3, or even 4 nm which explain the
tendency illustrated on figure 3. Consequently, our numerical approach accurately simulate
the experimental results of Bachelet and coll.\cite{bachelet} who demonstrate a enhanced growth in
height of the YSZ nanocrystals on a alpha-alumina substrate with planar defects. \\
\indent
We now turn to the evolution of the inter-plane spacings of the domains in the vertical
direction. \\
On figure 4 one can see the evolution of the number of domains with a given inter-plane
spacing in the vertical direction with regard to the computation time. It is clear that at the
end of the computation process the 10$^4$ domains reach the highest inter-plane spacing in the
vertical direction. This result is in perfect agreement with the experimental investigations of
Bachelet and coll.\cite{bachelet}. Indeed, these authors demonstrate that, on a $\alpha$-alumina substrate
with planar defects, the YSZ nanocrystals reach the [111] crystallographic direction in the
direction normal to the interface i.e. the vertical direction in our model.

\section{Conclusion}
We modelled the islanding, without deposition, of polycrystalline thin films by a Monte
Carlo process. The governing equation allowing to compute the energy of each of these
domains takes into account their height, their horizontal and vertical inter-plane spacings
with respect to the substrate. \\
\indent
Our numerical results traduce the experimental fact that the YSZ nanocrystals, which are
obtained after thermal treatment of an YSZ thin film on a alpha-alumina substrate with
planar defects, grow in height and reach the crystallographic direction [111] in the direction
normal to the interface. \\
\indent
The conclusion of this computation is that the influence of the planar defects of the substrate
can not be neglected as it leads to different crystallographic orientations of the nanocrys-
tals. The comprehension of the physical parameters involved in both the morphological and
epitaxial properties of YSZ nanocrystals on crystalline substrates is a key point to design
new systems suitable for photoluminescence applications.

\pagebreak

\begin{figure}
\includegraphics[width=17cm]{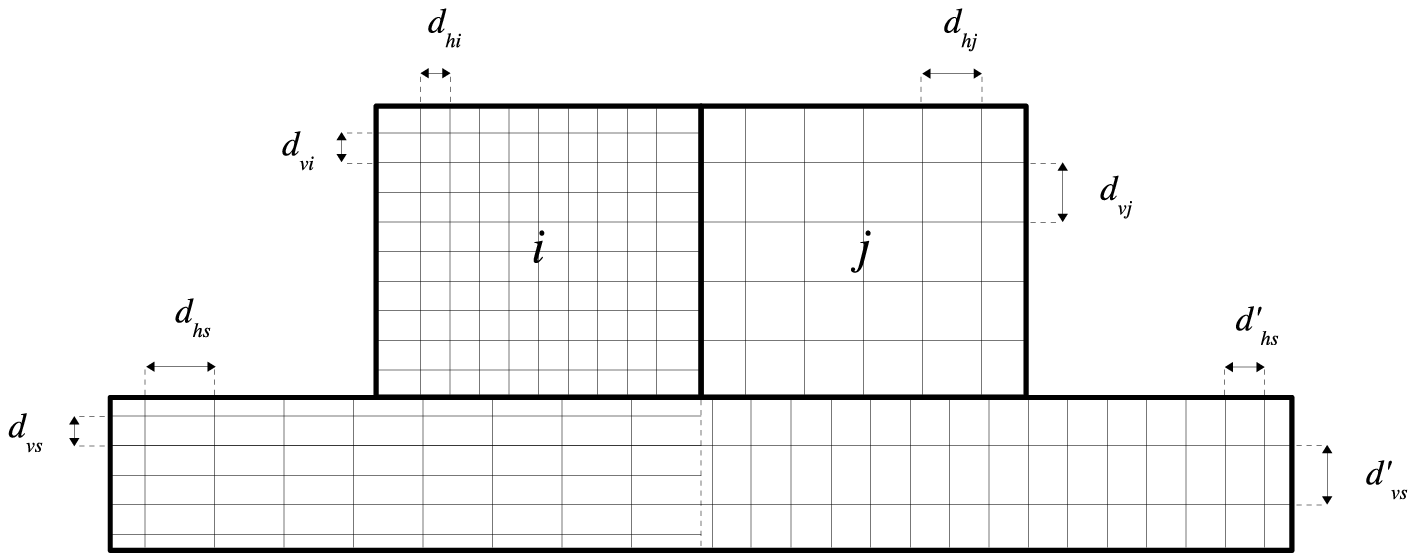}
\caption{Schematic representation of the interplane spacings of the domains and the substrate in the vertical and horizontal directions.}
\end{figure}

\begin{figure}
\includegraphics[width=17cm]{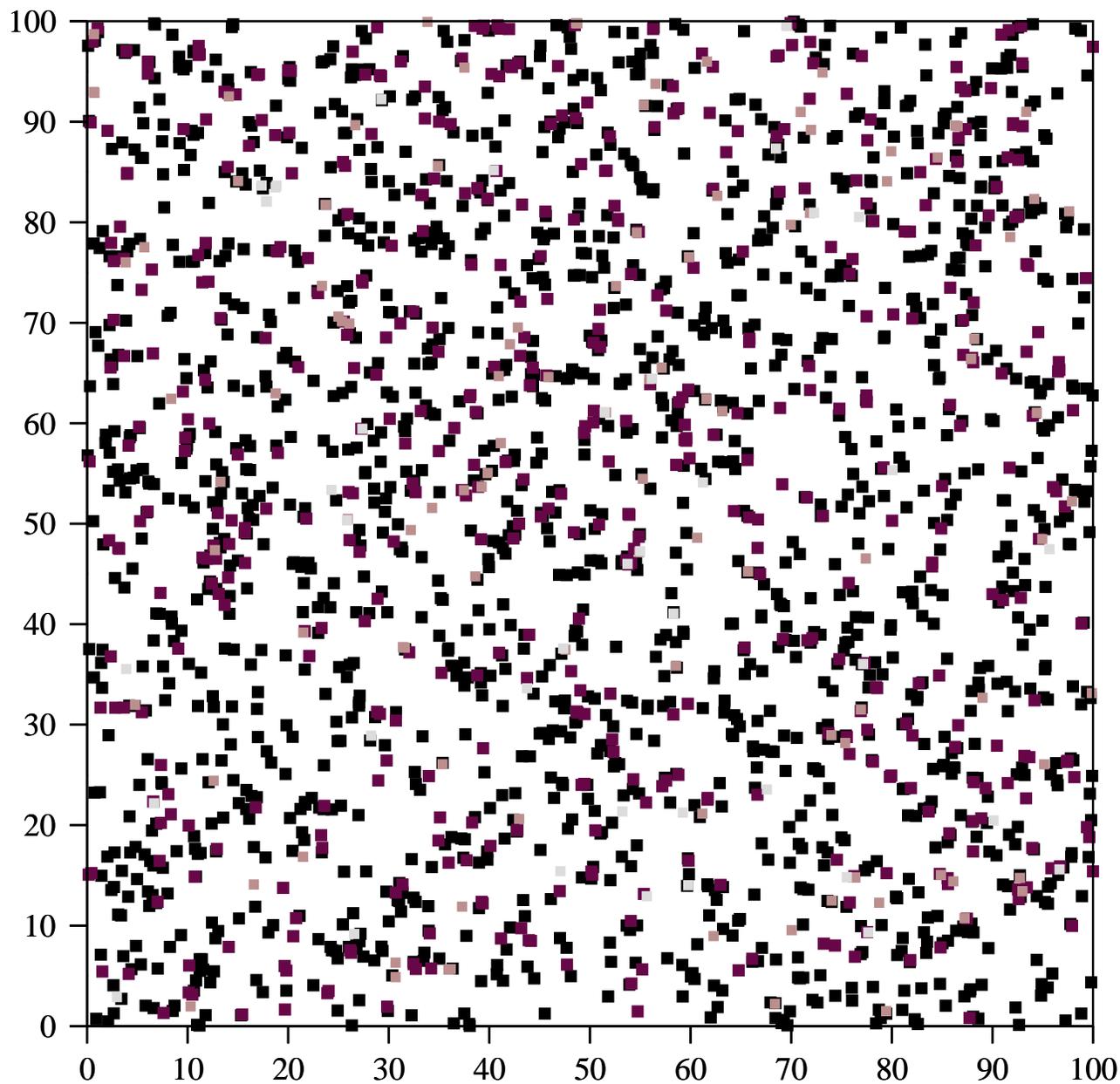}
\caption{Top view of the thin film after 10$^9$ MCS; the grey level of each domain is as dark as its height is small.}
\end{figure}

\begin{figure}
\includegraphics[width=10cm]{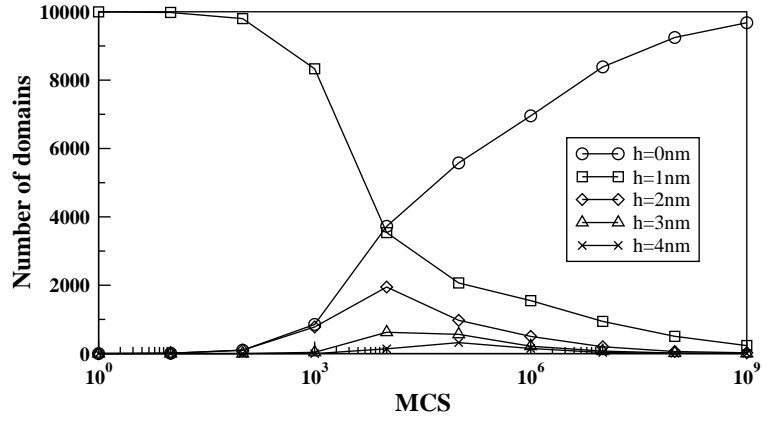}
\caption{Evolution of the heights of the domains as a function of MCS.}
\end{figure}

\begin{figure}
\includegraphics[width=10cm]{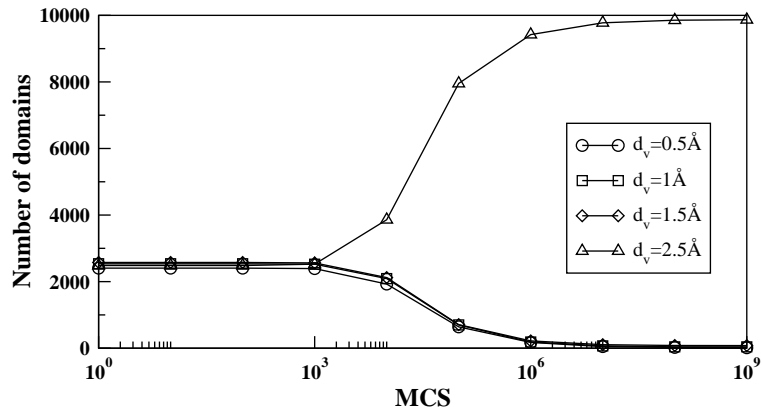}
\caption{Evolution of the interplane spacings of the domains in the vertical direction as a function of MCS.}
\end{figure}

\end{document}